\documentclass[twocolumn,showpacs,preprintnumbers,amsmath,amssymb]{revtex4}
\usepackage{graphicx}
\usepackage{dcolumn}
\usepackage{bm}
\begin{document}
\title{Phase transition within deformed Ising model}
\author{Alexander I. Olemskoi}
\email{olemskoi@ssu.sumy.ua}
\author{Olga V. Yushchenko}
\affiliation{Physical electronics department, Sumy State
University, 40007 Sumy, Ukraine}
\date{\today}
\begin{abstract}
Deformation of Ising Hamiltonian by means of replacing a site spin
$s_i$ by $s_i^q$ and statistics generalization with help of the
substituting deformed probability $p_i^q$ instead of $p_i$ are
studied jointly within mean--field scheme. Such deformed model is
shown to be related to the phase transition of the second order
with unusual set of critical indices depending essentially on the
deformation parameter $q$. Scaling relations turn out to be
invariant with respect to the deformation.
\end{abstract}
\maketitle
\section{Introduction}\label{sec:level1}

The Ising model is known to be a corner stone of the microscopic
theory of phase transitions \cite{a} --- \cite{d}. Evident
simplicity of this model is based on a supposition that every
lattice site $i$ has a spin taking only two magnitudes $s_i=\pm
1$. Although exact solution is not found for the three--dimension
Ising model, its using allows one to explain main peculiarities of
real phase transitions. Along this line, the qualitative picture
becomes clear already within framework of the mean--field
approach.

Recent considerations pay much attention to study complex systems
which self--similarity derives to generalization of the
Gibbs--Boltzmann statistics to Tsallis--type one (see Ref.\cite{2}
and references the\-re\-in). Formally, such a generalization is
performed by means of replacement of the probability $p_i$ by the
so called deformed probability $p_i^q$ with a positive index
$q\leq 1$. Here, we propose to complete such a procedure by the
relevant deformation on the microscopic level. Namely, we deform
the Ising Hamiltonian by means of replacing a site spin $s_i$ by
$s_i^q$. Such a deformation is shown to uphold the second order
 phase transition, whereas the set of critical indices becomes
depending on the parameter $q$ essentially.

The Letter is organized in the following manner. In Section 2
the $q$--deformed Hamiltonian is postulated and
simplified within the mean--field approach. Section 3
de\-als with determination of the fractional
average allowing us to obtain the definition of the order
parameter. In Section 4 $q$--deformed distribution
function is found to write a self--consistency equation for the
order parameter. Section 5 devotes to the
consideration of asymptotic solutions of the self--consistency
equation. The last Section 6 contains a brief
discussion concerning scaling relations.

\section{Main statements}\label{sec:level2}

We postulate the q--deformed Hamiltonian in the following form:
\begin{equation}
H=-\frac{1}{2}\sum_{i,j}{\mathcal J}_{ij}s_i^q s_j^q
-h\sum_i s_i^q.
\label{1}
\end{equation}
Here, summation runs over $N$ lattice sites $i\ne j$,
${\mathcal J}_{ij}$ is effective interaction potential,
$h$ is external field, $s_i=\pm 1$ is a site spin,
$q\leq 1$ is a deformation parameter.

In the simplest way, we use further the usual scheme of the mean--field
approximation  \cite{a} -- \cite{d}.
Within this approach, one follows to replace the multiplier $s_j^q$
by an average value $\langle s^q\rangle$.
Moreover, we shall take into account the only interaction of
nearest neighbor sites whose number equals $z$ and
potential is reduced to the constant ${\mathcal J}>0$.
Then, the mean--field effective Hamiltonian is as follows:
\begin{eqnarray}
H_{ef}=\sum_i\varepsilon_i;\quad
\varepsilon_i=-h_q s_i^q,\
h_q\equiv h + T_c\langle s^q\rangle
\label{2}
\end{eqnarray}
where a characteristic temperature $T_c\equiv z{\mathcal J}$ is introduced.

It is easy to foresee the postulated Hamiltonian will be meaningless
if the index $q\leq 1$ is not chosen to satisfy the condition
\begin{equation}
(-1)^q=-1.
\label{3}
\end{equation}
Then, the complex representation $-1\equiv {\rm e}^{i\pi(2n+1)}$,
$n=0,\pm 1,\dots$ arrives at the set of rational numbers
\begin{equation}
q=\frac{2m+1}{2n+1},\quad m\leq n
\label{4}
\end{equation}
with integers $m,n=0,\pm 1,\dots$. Obviously, above reduction
of the continuous set $q\in[0,1]$ into the manifold of rational
numbers (\ref{4}) is caused by the discrete symmetry of the Ising model
being invariant with respect to the transformations
$\{s_i\}\to\{-s_i\}$, $h\to -h$.

As a result, the site energy in the effective Hamiltonian (\ref{2})
takes the linear form
\begin{equation}
\varepsilon_i=-h_q s_i.
\label{2a}
\end{equation}

\section{Calculation of the fractional average}\label{sec:level3}

Now, we should express the fractional average $\langle s^q\rangle$
by means of the order parameter $\eta\equiv\langle s\rangle$.
Obviously, such a problem will be quite correct only for self--similar systems
where it is reduced to definition of an index $p(q)$ in the relation
\cite{O}
\begin{equation}
\langle s^q\rangle\equiv\eta^{p(q)}.
\label{9}
\end{equation}

If relevant distribution varies very slightly near the origin $x=0$
and decays abruptly in the limit $x\to\infty$ (see Eqs.(\ref{11}), (\ref{12})
below), we have the following estimations:
\begin{eqnarray}
\langle x^q\rangle\sim\int
x^q{\rm d}x\sim x^{1+q},\quad
\langle x\rangle\sim\int x{\rm d}x\sim x^2.
\end{eqnarray}
These derive immediately to the principle relation
\begin{equation} \langle
x^{q}\rangle\sim\langle x\rangle^{p(q)},\quad p(q)=\frac{1+q}{2}.
\label{8}
\end{equation}
Respectively, the effective field in Eq.(\ref{2a})
takes the form
\begin{equation}
h_q=h+T_c\eta^{\frac{1+q}{2}}
\label{10}
\end{equation}
that differs from the usual one with change of the order parameter
$\eta$ by the power--law function $\eta^{\frac{1+q}{2}}$.

\section{Statistical scheme}\label{sec:level4}

Along the line of using $q$--deformed quantities, we need
to state on the Renyi entropy
\begin{equation}
S_q\equiv(1-q)^{-1}\ln\sum\limits_{\{s_i\}} P^q\{s_i\}
\label{0}
\end{equation}
which is easy seen to take the Boltzmann form in the limit
$q\to 1$. Then, the maximum entropy principle taken with
accounting the normalization condition \break $\sum_{\{s_i\}}
P\{s_i\}=1$ and the definition of the $q$--deformed internal
energy $E_q\equiv\sum_{\{s_i\}}\varepsilon_i(s_i)P^q\{s_i\}$
derives to the distribution function
\begin{eqnarray}
&P\{s_i\}=\prod\limits_i p_i(s_i),\ \
p_i=z_q^{-1}\exp_q\left(-\beta\langle 1\rangle_q
\varepsilon_i\right),&
\label{11}\\
&\langle 1\rangle_q\equiv\sum\limits_{\{s_i\}}p_i^q(s_i)&
\label{11a}
\end{eqnarray}
where $z_q$ is the specific partition function related to one
site, $\beta\equiv \frac{1}{T}$ is inverse temperature measured in
the energy units; $q$--deformed exponential is as follows:
\begin{equation}
\exp_q(x)\equiv\left\{\begin{array}{ll}
\left[1+(1-q)x\right]^{\frac{1}{1-q}}\ &{\rm at}\
1+(1-q)x>0,\\
0 &{\rm otherwise}.
\end{array} \right.
\label{12}
\end{equation}
Such a form is known to generalize usual exponential related to
the limit $q\to 1$. It allows one to generate in usual manner the
set of $q$--deformed hyperbolic functions: $\sinh_q(x)$,
$\cosh_q(x)$, $\tanh_q(x)$ and $\coth_q(x)$ \cite{2}.

We calculate now the site partition function
\begin{equation}
z_q\equiv\sum_{s_i=\pm 1}\exp_q
\left(-\beta\langle 1\rangle_q
\varepsilon_i\right)
=2\cosh_q(\beta\langle 1\rangle_q h_q)
\label{14}
\end{equation}
that differs from the usual one with $q$--deformation and
$\langle 1\rangle_q$ factor appearance. According to Eqs.(\ref{11}),
(\ref{11a}), (\ref{14}) the latter is determined by the equation
\begin{equation}
\langle 1\rangle_q=\frac
{\left[\exp_q\left(\beta\langle 1\rangle_q h_q\right)\right]^q+
\left[\exp_q\left(-\beta\langle 1\rangle_q h_q\right)\right]^q}
{\left[\exp_q\left(\beta\langle 1\rangle_q h_q\right)+
\exp_q\left(-\beta\langle 1\rangle_q h_q\right)\right]^q}.
\label{14a}
\end{equation}
With accounting Eq.(\ref{14}), the order parameter
$\eta\equiv\langle s\rangle$ is defined through the
self--consistency equation
\begin{equation}
\eta^{\frac{1+q}{2}}=\tanh_q
[\beta\langle 1\rangle_q h_q(\eta)]
\label{15}
\end{equation}
following from Eqs.(\ref{9}), (\ref{11}), (\ref{12})
and condition (\ref{3}).

Solutions of the system (\ref{14a}), (\ref{15}) are shown in Fig.1
for the external field $h=0$ and different values of index $q$.
Main peculiarity of the temperature dependencies $\eta(T)$,
$\langle 1\rangle_q(T)$ is in decreasing the order parameter
$0\leq\eta\leq 1$ accompanied by the relevant increasing the
parameter $1\leq\langle 1\rangle_q\leq 2$ at the temperature
growth within the domain
\begin{equation}
T_{cq}\leq T\leq T_q,\quad T_{cq}\equiv(1-q)T_c,\ \ T_q\equiv 2^{1-q}T_c
\label{13}
\end{equation}
(at $T<T_{cq}$ one has $\eta=\langle 1\rangle_q=1$, whereas at
$T>T_q$ there are $\eta=0$, $\langle 1\rangle_q=2$). Thus,
decreasing the deformation parameter $1\geq q\geq 0$ derives to
monotonous growth of the order parameter that transforms smoothly
falling down dependence $\eta(T)$ into step--like one.

\vspace{1cm}

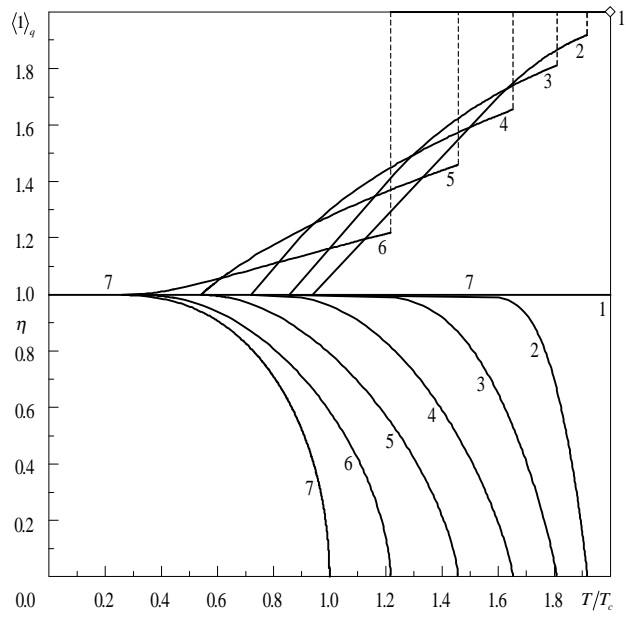
Fig.1. Temperature dependencies of both the renormalization
parameter (upper panel) and the order parameter (low panel) for
different indexes $q$ (curves 1 --- 7 relate to $q=0,\ 1/17,\
1/7,\ 3/11,\ 5/11,\ 5/7,\ 1$, respectively).
\vspace{1cm}

As demonstrated in Fig.2, the effect of the external field is
similar to the usual case $q=1$: high--temperature dependence
$\eta(h)$ has the monotonically growing form (dashed curves) that
takes, at the low temperatures $T<T_q$, the falling down domain
located in the vicinity of the point $h=\eta=0$ (solid curves). At
fixed temperatures, decrease of the deformation parameter is seen
to arrive at the more sharply defined non--monotone dependence. At
critical field $h_q=-T_c(1-T/T_{cq})$, the dependence $\eta(h)$
undergoes a break, whereas in the limit of the small fields $h\ll
h_q$ one has
\begin{equation}
\eta^{\frac{1+q}{2}}\simeq -\frac{h/T_c}{1-T/T_q}.
\label{13a}
\end{equation}

\vspace{2cm}
Fig.2. Field dependencies of the order parameter: dashed curves 1
--- 6 correspond to the fixed index $q=5/11$ and temperatures
$T/T_c= 1.5,\ 1.7,\ 1.9,\ 2.2,\ 2.5,\ 2.8$; solid curves 1 --- 6
relate to indexes $q=0,\ 1/7,\ 3/11,\ 5/11,\ 11/17,\ 1$ at $T=0.5
T_c$. Insertion shows details of the coordinate origin
neighborhood.

\section{Asymptotic regimes}\label{sec:level5}

To obtain analytical description, we look over the possible
asymptotic solutions of the system (\ref{14a}), (\ref{15})
in limiting cases of both zero and non--zero external fields.

\subsection*{Zero field}

For small order parameters $(\eta\ll 1)$, one has the
expansions
\begin{eqnarray}
&\tanh_q(x)\simeq x-\frac{q(2-q)}{3}x^3,&
\label{16}\\
&\langle 1\rangle_q\simeq 2^{1-q}\left[1+
2^{2(1-q)}(1-q)^2\left(\beta h_q\right)^2\right].&
\end{eqnarray}
Then, the temperature dependence of the order parameter takes the
power--law form
\begin{eqnarray}
&\eta\simeq A^\beta\left(1-\frac{T}{T_q}\right)^\beta,&
\label{17}\\
&A\equiv\frac{3}{(3-2q)(2q-1)},\
\beta={\frac{1}{1+q}},\ T_q\equiv 2^{1-q}T_c&
\label{17a}
\end{eqnarray}
that corresponds to critical domains $T_q-T\ll T_q$ in Fig.1.
At $q<1$, the index $\beta$ exceeds the usual magnitude $\beta=1/2$.

At the marginal magnitude $q=0$, one has
\begin{equation}
\tanh_0(x)=x,\quad \langle 1\rangle_0=2
\label{18}
\end{equation}
and equation (\ref{15}) arrives at the condition
\begin{equation}
\sqrt{\eta}=\frac{T_q}{T}\sqrt{\eta}.
\label{19}
\end{equation}
It remains valid for arbitrary values $\eta$ at $T=T_q$.
So, this case corresponds to the step--like curve 1 in Fig.1.

In the limit $q\ll 1$, there are the expansions
\begin{equation}
\langle 1\rangle_q\simeq 2^{1-q},\quad
\tanh_q(x)\simeq x\left(1-2q~\frac{x^2}{1-x^2}\right).
\label{20}
\end{equation}
Respectively, the order parameter \begin{equation}
\eta\simeq \left(1+\frac{2q}{1-T/T_q}\right)^{-1}
\label{21}
\end{equation}
decays very fast within tight window $1-T/T_q\sim q\ll
1$ (see curve 2 in Fig.1).

Finally, we consider the form of the temperature dependence of the
free energy $F\equiv -TN\ln(z_q)$ near the critical temperature
$T_q$. Here, the hyperbolic cosine in Eq.(\ref{14}) can be
expanded over the argument $\beta\langle 1\rangle_q h_q\simeq
\beta T_q\eta^{\frac{1+q}{2}}$
in series comprising of even terms only. In accordance with
relevant temperature dependence (\ref{17}), these terms have
orders $[(T_q-T)/T_q]^n$ with the lowest index $n=2$ because the
first--order term $(n=1)$ is suppressed by self--action effects.
Thus, in the limit $(T_q-T)/T_q\ll 1$ we obtain
$F\sim T_q[(T_q-T)/T_q]^{2}$ and the specific heat $C\sim{\rm d}^2
F/{\rm d}T^2$ does not vary with temperature. This results in
the magnitude
\begin{equation}
\alpha=0
\label{23a}
\end{equation}
of the critical index of the dependence
$C\sim[(T_q-T)/T_q]^{-\alpha}$.

\subsection*{Non--zero field}

Here, we start with the approximate equation
\begin{eqnarray}
\eta^{\frac{1+q}{2}}\simeq 2^{1-q}\beta h_q\left[1-\frac{2^{2(1-q)}}{A}
(\beta h_q)^2\right]
\label{24}
\end{eqnarray}
following from Eqs.(\ref{15}), (\ref{16})
(notations $h_q$ and $A$ are given with Eqs.(\ref{10}), (\ref{17a})).
In the limit $h\to 0$, differentiation of Eq.(\ref{24}) with respect
to the field $h$ yields the following expression for the susceptibility
$\chi\equiv{\rm d}\eta/{\rm d}h$:
\begin{equation}
T_c\chi=2(1+q)^{-1}\eta^{\frac{1-q}{2}}
\left(\frac{T-T_q}{T_q}+\frac{3}{A}\eta^{1+q}\right)^{-1}.
\label{25}
\end{equation}

Thus, in disordered state $(\eta=0,~T>T_q)$ one has the
susceptibility $\chi=0$. Physically, this means a suppression of
critical fluctuations in the disordered phase --- quite
differently from the usual picture.

In ordered phase $(\eta\ne 0,~T<T_q)$, inserting Eq.(\ref{17})
into Eq.(\ref{25}) we get the power--law dependence
\begin{equation}
T_c\chi=\frac{A^{\frac{1}{2}\frac{1-q}{1+q}}}{1+q}
\left(\frac{T_q-T}{T_q}\right)^{-\gamma}
\label{26}
\end{equation}
with the critical index
\begin{equation}
\gamma=1-\frac{1}{2}~\frac{1-q}{1+q}.
\label{27}
\end{equation}
With decreasing the deformation parameter $1\geq q\geq 0$, this
index falls down monotonically from the usual value $\gamma=1$ to
$\gamma=1/2$.

In the opposite limit $h\to\infty$, Eq.(\ref{24}) related to the
critical point $T=T_q$ derives to the power--law dependence
\begin{equation}
\eta\simeq\left(A~\frac{h}{T_c}\right)^{\frac{1}{\delta}}
\label{29}
\end{equation}
with the critical index
\begin{equation}
\delta=\frac{3}{2}(1+q).
\label{30}
\end{equation}

\section{Discussion}\label{sec:level6}

Above mean--field consideration shows the deformation of Ising
Hamiltonian by means of replacing a site spin $s_i$ by a power
function $s_i^q$ derives to the phase transition of the second
order with unusual set of critical indices (\ref{23a}),
(\ref{17a}), (\ref{27}) and (\ref{30}). The zeroth magnitude of
the first of them is caused apparently by the mean--field
approach. It is easy to convince also in zeroing indexes
$\varepsilon$, $\zeta$ defined by the field dependence of the
specific heat $C\sim h^{-\varepsilon}$ and the space dependence of
the correlation function $G(r)\sim r^{-(d-2+\zeta)}$ \cite{L}.
Moreover, there are critical indexes $\mu$ and $\nu$ defining the
field--temperature dependencies $\xi\sim h^{-\mu}$ and $\xi\sim
(T_q-T)^{-\nu}$ of the correlation length $\xi$. Among complete set
of these indexes, only two are independent, whereas the rest are
known to be determined by the scaling relations. Then, the question
of fundamental importance arises how influences the deformation on
these relations.

The first of them, Widom relation, takes the usual form \cite{L}
\begin{equation}
\beta\delta=\beta+\gamma
\label{31}
\end{equation}
that follows from comparison of two chains of definitions:
$\eta\sim\chi h\Rightarrow h\sim (T_q-T)^{\beta+\gamma}$ and
$h\sim\eta^\delta\sim(T_q-T)^{\beta\delta}$.
Using the relation $\xi\sim h^{-\mu}$ in the latter, one has
\begin{equation}
\mu(\beta+\gamma)=\nu.
\label{32}
\end{equation}
Combination of Eqs.(\ref{31}), (\ref{32}) yields the relation
$\beta\delta\mu=\nu$ that together with Eqs.(\ref{17a}) and
(\ref{30}) gets $3\mu=2\nu$.
Finally, using the fluctuation--dissipation theorem derives to
the equality \cite{L}
\begin{equation}
2\nu=\gamma
\label{33}
\end{equation}
that defines the rest of indexes:
\begin{eqnarray}
\mu=\frac{1}{3}\left(1-\frac{1}{2}~\frac{1-q}{1+q}\right),\quad
\nu=\frac{1}{2}\left(1-\frac{1}{2}~\frac{1-q}{1+q}\right).
\label{35}
\end{eqnarray}

What about the Essam--Fisher equality, it takes the form
\begin{equation}
\alpha+\beta+\gamma=\frac{3}{2}
\label{32a}
\end{equation}
following from the consequence
$C(T_q-T)^2\sim\eta^{\frac{1+q}{2}}h \Rightarrow h\sim
(T_q-T)^{\frac{3}{2}-\alpha}$ that differs from the usual one by
replacing $\eta$ by $\eta^{\frac{1+q}{2}}$.

It is worthwhile to note all of equalities
(\ref{31}) --- (\ref{33}), (\ref{32a}) are invariant with respect to
the deformation.

In this work, financial support by STCU, project 1976,
is gratefully acknowledged.

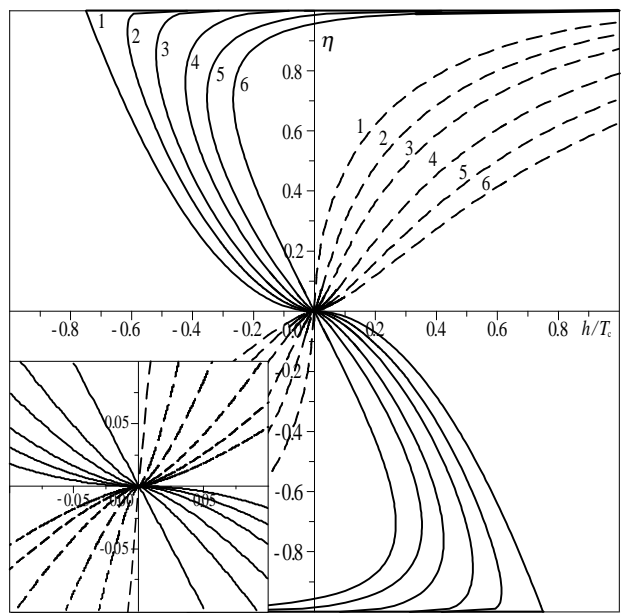
\end{document}